\DeclareSymbolFont{UPM}{U}{eur}{m}{n}
\DeclareMathSymbol{\umu}{0}{UPM}{"16}
\let\oldumu=\umu
\renewcommand\umu{\ifmmode\oldumu\else\math{\oldumu}\fi}

\newcounter{magicrownumbers}
\newcommand\rownumber{\stepcounter{magicrownumbers}\arabic{magicrownumbers}}

\documentclass[twocolumn]{aastex63}

\usepackage{longtable}
\usepackage{chngcntr}

\usepackage{hyperref}

\expandafter\def\expandafter\UrlBreaks\expandafter{\UrlBreaks
  \do\*\do\-\do\~\do\'\do\"\do\-}%

\received{March 4, 2020}
\revised{January 22, 2021}
\accepted{---}
\submitjournal{PSJ}

\usepackage{todonotes}

\shorttitle{Accurate Machine Learning Atmospheric Retrieval}
\shortauthors{Himes et al.}
\graphicspath{{./}{figures/}}

\begin{document}

\title{Accurate Machine Learning Atmospheric Retrieval via a Neural Network Surrogate Model for Radiative Transfer}

\correspondingauthor{Michael D. Himes}
\email{mhimes@knights.ucf.edu}

\author[0000-0002-9338-8600]{Michael D. Himes}
\affiliation{Planetary Sciences Group, Department of Physics, University of Central Florida}

\author[0000-0002-8955-8531]{Joseph Harrington}
\affiliation{Planetary Sciences Group, Department of Physics and Florida Space Institute, University of Central Florida}


\author[0000-0003-2868-6983]{Adam D. Cobb}
\affil{Department of Engineering Science, University of Oxford}

\author[0000-0001-9854-8100]{At{\i}l{\i}m G\"{u}ne\c{s} Bayd{\rlap{\.}\i}n}
\affiliation{Department of Engineering Science, University of Oxford}

\author[0000-0001-8185-6094]{Frank Soboczenski}
\affiliation{SPHES, King's College London}

\author[0000-0001-9011-4420]{Molly D. O'Beirne}
\affiliation{Department of Geology and Environmental Science, University of Pittsburgh}

\author[0000-0003-0550-3224]{Simone Zorzan}
\affil{ERIN Department, Luxembourg Institute of Science and Technology}

\author[0000-0003-1562-4679]{David C. Wright}
\affiliation{Planetary Sciences Group, Department of Physics, University of Central Florida}

\author[0000-0002-1177-113X]{Zacchaeus Scheffer}
\affiliation{Planetary Sciences Group, Department of Physics, University of Central Florida}

\author[0000-0003-0354-9325]{Shawn D. Domagal-Goldman}
\affiliation{NASA Goddard Space Flight Center, Greenbelt, MD}

\author[0000-0001-6285-267X]{Giada N. Arney}
\affiliation{NASA Goddard Space Flight Center, Greenbelt, MD}

\begin{abstract}

Atmospheric retrieval determines the properties of an atmosphere based on its measured spectrum. 
The low signal-to-noise ratios of exoplanet observations require a Bayesian approach to determine posterior probability distributions of each model parameter, given observed spectra. 
This inference is computationally expensive, as it requires many executions of a costly radiative transfer (RT) simulation for each set of sampled model parameters. 
Machine learning (ML) has recently been shown to provide a significant reduction in runtime for retrievals, mainly by training inverse ML models that predict parameter distributions, given observed spectra, albeit with reduced posterior accuracy. 
Here we present a novel approach to retrieval by training a forward ML surrogate model that predicts spectra given model parameters, providing a fast approximate RT simulation that can be used in a conventional Bayesian retrieval framework without significant loss of accuracy. 
We demonstrate our method on the emission spectrum of HD 189733 b and find good agreement with a traditional retrieval from the Bayesian Atmospheric Radiative Transfer (BART) code (Bhattacharyya coefficients of 0.9843--0.9972, with a mean of 0.9925, between 1D marginalized posteriors).
This accuracy comes while still offering significant speed enhancements over traditional RT, albeit not as much as ML methods with lower posterior accuracy.
Our method is $\sim9\times$ faster per parallel chain than BART when run on an AMD EPYC 7402P central processing unit (CPU). 
Neural-network computation using an NVIDIA Titan Xp graphics processing unit is 90--180$\times$ faster per chain than BART on that CPU.
\end{abstract}

\keywords{techniques: retrieval --- techniques: machine learning --- methods: statistical --- planets and satellites: atmospheres --- planets and satellites: individual (HD 189733 b)}


\section{Introduction}
\label{sec:intro}

Over the past decades, exoplanet studies have expanded from their detection to include characterization of their atmospheres via retrieval \citep[see reviews by][]{SeagerDeming2010araaExoplanetAtmospheres, DemingSeager2017jgreExoplanetAtmospheresReview}. 
Retrieval is the inverse modeling technique whereby forward models of a planet's spectrum are compared to observational data in order to constrain the model parameters \citep{Madhusudhan2018bookRetrieval}. 
These typically include the shape of the thermal profile, abundances of species, and condensate properties. While some solar system objects can be characterized with simpler approaches (such as Levenberg--Marquardt minimization) due to their high signal-to-noise ratios \citep[e.g.,][]{KoskinenEtal2016grlSaturnBenzeneRetrieval}, retrieval on noisy exoplanet spectra require Bayesian methods to provide a distribution of models that can explain the observed data.  
The posterior distribution resulting from a Bayesian retrieval places limits on each model parameter (within some range, an upper or lower limit, or equally probable for all values considered), informing the statistical significance of the result.  

Bayesian retrieval methods involve evaluating thousands to millions of spectra, integrating over the observational bandpasses, and comparing to observations.  
Depending on model complexity, this requires hundreds to thousands of parallelizable compute hours, resulting in hours to days of runtime.  
Calculating the model spectra by solving the radiative transfer (RT) equation takes the vast majority of compute time. 

Machine learning (ML) encompasses algorithms that learn representations of and uncover relationships within a collection of data samples. 
Deep learning \citep{GoodfellowEtal2016bookDeepLearning} is a subfield of ML that is based on neural networks, which are highly flexible differentiable functions that can be fit to data. 
Neural networks can classify images \citep[e.g.,][]{KrizhevskyEtal2012nipsAlexNet, Simonyan15, SzegedyEtal2015cvprGoingDeeper, he2016deep, HuangEtal2017cvprDenselyConnectedConvNets}, recognize speech \citep[e.g.,][]{ChorowskiEtal2014nipsContinuousSpeechRecognition, amodei2016deep, ChanEtal2016icasspSpeechRecognition, XiongEtal2016ieeeHumanParitySpeechRecognition}, and translate between languages \citep[e.g.,][]{ChoEtal2014StatisticalMachineTranslation, BahdanauEtal2015iclrNeuralMachineTranslation, RanzatoEtal2016iclrRNNMachineTranslation, SennrichEtal2016MachineTranslationRareWords, wu2016google}. 
Neural networks consist of a hierarchy of layers that contain nodes performing weighted (non)linear transformations of their inputs, through a series of hidden layers, to the desired output. 
For example, for a retrieval, one might have the input layer receive the observed spectrum, hidden layers extract features, and the output layer predict the underlying atmospheric parameters.
Neural-network training conventionally uses gradient-based optimization, iteratively adjusting the weights of the connections between nodes to minimize the error between the neural network's prediction and the desired output \citep{RumelhartEtal1986natureBackPropagation}.

Recent applications of ML to atmospheric retrieval reduced compute time from hundreds of hours to minutes or less.  
\citet{MarquezNeilaEtal2018natasHELA} presented a random forest of regression trees to build predictive distributions comparable to the posterior distributions of traditional Bayesian retrievals.  
\citet{ZingalesWaldmann2018ajExoGAN} utilized a generative adversarial network \citep[GAN;][]{Goodfellow2014nipsGAN} to retrieve distributions for model parameters. 
\citet{WaldmannGriffith2019natasPlanetNet} used a convolutional neural network (CNN) to map spatial and spectral features across Saturn.
In \citet{CobbEtal2019ajPlanNet}, we introduced \texttt{plan-net}, an ensemble of Bayesian neural networks that uses parameter correlations to inform the uncertainty on retrieved parameters.  
\citet{HayesEtal2019arxivUnsupervisedPCA} demonstrated a new approach to ML retrieval by applying \textit{k}-means clustering to a principal component analysis of the observed spectrum to inform a standard Bayesian retrieval.  
\citet{JohnsenMarley2019arxivMLP} showed that a dense neural network can provide quick estimations of atmospheric properties.

While these approaches are promising, all except \citet{HayesEtal2019arxivUnsupervisedPCA} suffer from a common deficiency: the reduction in computational time is accompanied by a reduction in posterior accuracy because they make significant approximations when performing Bayesian inference.
For ML to become an integral part of atmospheric retrieval, the accuracy of the posterior approximation must be preserved.

The solution lies in simulation-based inference methods \citep{cranmer2019frontier}. 
While directly using a simulator (e.g., RT code) requires a consistent amount of compute time for each new inference (e.g., retrieval), surrogate models that emulate the simulator (e.g, neural networks) allow new data to be quickly evaluated after an upfront computational cost to train the surrogate \citep{kasim2020up, munk2019deep}. 
ML- and simulation-based inference approaches have been successfully applied to a variety of tasks ranging from quantum chemistry \citep{GilmerEtal2017icmlQuantumChemistry} to particle physics \citep{brehmer2018guide, BaydinEtal2019neuripsQuestForPhysics}, resulting in significant reductions in compute cost with minimal loss in accuracy.
Similar approaches have been used by the Earth science community to reduce the computational burden of forward modeling of spectra, retrieval of surface conditions, and atmospheric correction \citep[e.g.,][]{Atzberger2004rseBiophysicalCanopyRetrievalML, GarciaCuestaEtal2009springerMLRT, RiveraEtal2015remsensEmulatorToolboxRTM, VerrelstEtal2015igarssSurrogateRTM, GomezDansEtal2016remsensGaussianProcessesRTM, VerrelstEtal2016grslBiophysicalRetrieval, VerrelstEtal2017remsensReflectanceSpectraEmulation, ChernetskiyEtal2018asrHyperspectralSimulation, YinEtal2018igarssAtmosphericCorrection, VicentEtal2018EmulationVsInterpolationRT, BueEtal2019amtNeuralNetworkRT}.

\begin{figure}[tb]
\centering
\includegraphics[width=0.47\textwidth, clip]{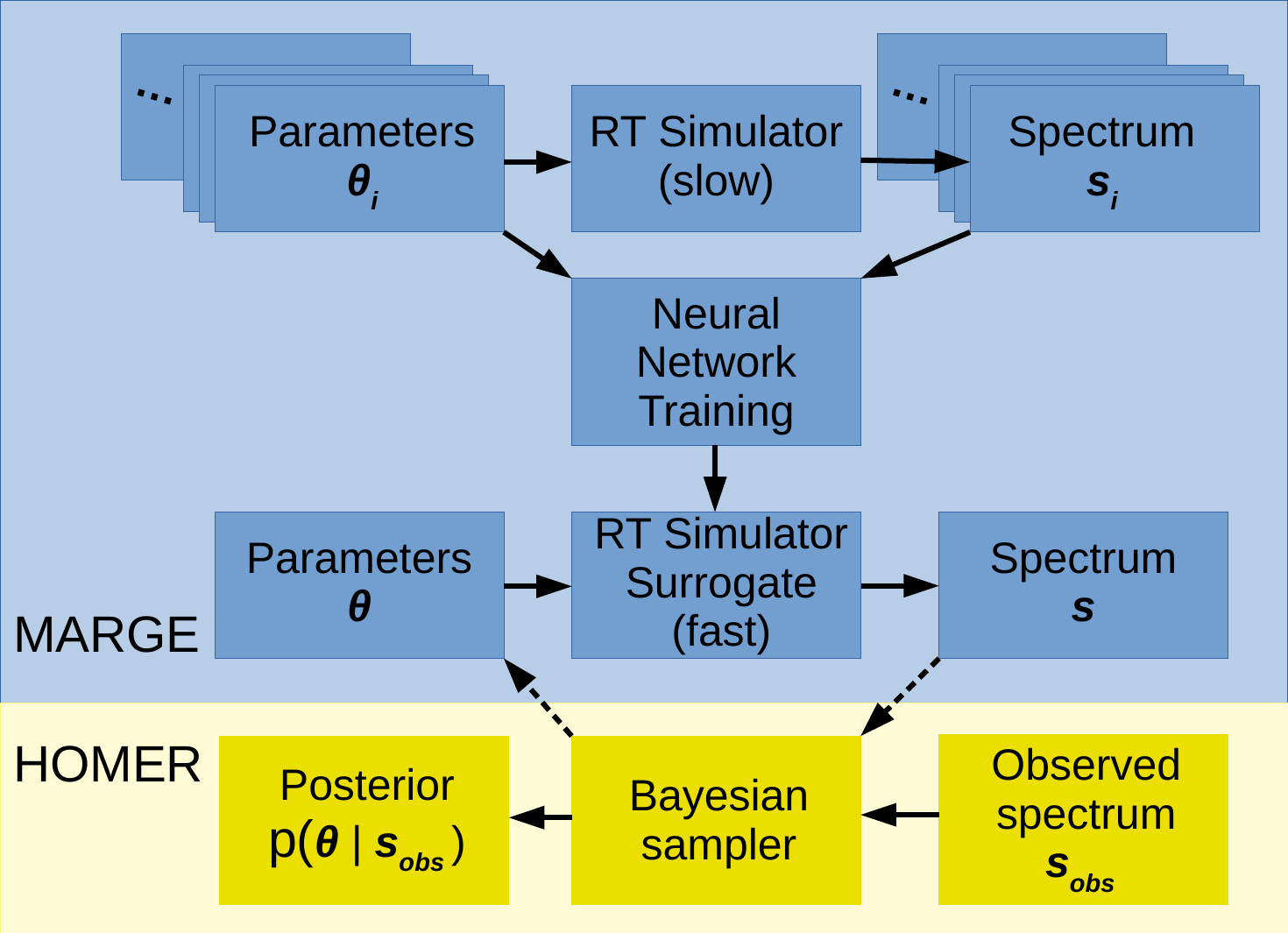}
\caption{Schematic diagram of our inverse modeling method, color-coded based on the scope of our software packages.  MARGE (Section \ref{sec:marge}) generates a data set based on a deterministic, forward process (e.g., RT) and trains a surrogate model to approximate that process.  Using the trained surrogate, HOMER (Section \ref{sec:homer}) infers the inverse process (e.g., atmospheric retrieval) by simulating many forward models and comparing them to the target data (e.g., an observed spectrum) in a Bayesian framework.}
\label{fig:schematic}
\end{figure}

Here we present a novel application of this approach to retrieval, which uses a neural-network model of RT within a Bayesian framework, apply it to the emission spectrum of HD 189733 b, and compare the results to a classical retrieval using the RT code that trained the surrogate model.  
Our general method is to (1) generate a data set over some parameter space, (2) train a surrogate forward model on the generated data, and (3) infer the inverse process via a Bayesian sampler (Figure \ref{fig:schematic}).
Our approach circumvents the existing limitations of ML retrieval methods, which seek to directly learn the inverse process, by learning the forward, deterministic process (RT) and using the simulator surrogate in a standard inference pipeline.  
This approach preserves the accuracy of the Bayesian inference and, while slower than direct ML retrieval, is still much faster than computing RT.

In Section \ref{sec:methods}, we describe our approach in detail as well as introduce the software packages that implement the method. 
Section \ref{sec:results} discusses the results.
Finally, Section \ref{sec:conclusions} presents conclusions. 

\section{Methods}
\label{sec:methods}

\subsection{Model Training}

To train a neural network for our approach (Figure \ref{fig:schematic}), we generate a data set of spectra using the Bayesian Atmospheric Radiative Transfer (BART) code \citep{HarringtonEtal2020apjsBART1, CubillosEtal2020apjsBART2, BlecicEtal2020apjsBART3}.

The atmospheric models consist of 100 log-uniform layers spanning pressures from $10^{-8}$ to 100 bar, and we assume that the planet radius corresponds to a pressure of 0.1 bar.
We use the five-parameter temperature--pressure profile, $T(p)$, parameterization of \citet{LineEtal2013apjRetrieval1}: $\kappa$, the Planck mean infrared opacity; $\gamma_1$ and $\gamma_2$, the ratios of the Planck mean visible and infrared opacities for each of two streams; $\alpha$, which controls the contribution of the two streams; and $\beta$, which represents albedo, emissivity, and energy recirculation. 
We allow the radius ($R_p$), mass ($M_p$), and semimajor axis ($a$, adjusts the temperature at the top of the atmosphere due to stellar irradiation) of the planet to vary to encompass a range of hot Jupiters.  
We also include a free parameter for each of the uniform vertical abundance profiles of H$_2$O, CO$_2$, CO, and CH$_4$.

We allow a wide range of values without regard for physical plausibility, except by enforcing that (1) the H$_2$/He ratio remains constant, (2) the total relative abundances of molecules in the atmosphere equals 1, and (3) the $T(p)$ profile does not exceed the temperature range of the line lists.  
For example, this could lead to models with H$_2$O at conditions where it dissociates \citep{ArcangeliEtal2018apjWASP-18bNegH}, though in the case of HD 189733 b, such models would be rejected with a high probability due to a poor fit.
We note that these are not fundamental constraints of our approach; other constraints (e.g., enforcing thermochemical equilibrium, keeping elemental ratios within some range) may be used when generating the data set to train the surrogate model.

\begin{table}[tb]
\caption{Forward Model Parameter Space}\vskip -.1in
\label{tbl:hd189paramspace}
\begin{center}
\begin{tabular}{ l  c c }
 \toprule
Parameter & Minimum & Maximum\\
\hline
$\textrm{log}\ \kappa$                   &  -5.0 &  1.0 \\
$\textrm{log}\ \gamma_1$                 &  -2.0 &  2.0 \\
$\textrm{log}\ \gamma_2$                 &  -1.3 &  1.3 \\
$\alpha$                                 &   0.0 &  1.0 \\
$\beta$                                  &   0.7 &  1.3 \\
$R_p$ ($R_{J}$)                          &   0.8 &  1.5 \\
$M_p$ ($M_{J}$)                          &   0.8 &  1.5 \\
$a$ (AU)                                 &   0.2 &  0.4 \\
$\textrm{log}\ \textrm{H}_2\textrm{O}$   & -13   & -0.5  \\
$\textrm{log}\ \textrm{CO}_2$            & -13   & -0.5  \\
$\textrm{log}\ \textrm{CO}$              & -13   & -0.5  \\
$\textrm{log}\ \textrm{CH}_4$            & -13   & -0.5  \\
\hline
\end{tabular}
\end{center}
\end{table}

For opacities, we use HITEMP for H$_2$O, CO, and CO$_2$ \citep{Goorvitch1994apjsHiTempCO, TashkunEtal2003jqsrtHiTempCO2, BarberEtal2006mnrasHiTempH2O, RothmanEtal2010jqsrtHITEMP}, HITRAN for CH$_4$ \citep{NiedererEtal2008cijcInfraredCH4, BoudonEtal2010jqsrtCH4FarIR, NikitinEtal2010jqsrtCH4GOSAT, NikitinEtal2011jms2micronCH4, BrownEtal2013jqsrtMethaneHITRAN2012, CampargueEtal2013jmsPlanetaryCH4, DaumontEtal2013jqsrt2micronCH4, NiedererEtal2013jmsRotovibCH4, NikitinEtal20132halfmicronCH4, Rothman2013jqsrtHITRAN2012}, and collision-induced absorptions of H$_2$--H$_2$ and H$_2$--He \citep{BorysowEtal2001jqsrtH2H2highT, Borysow2002aapH2H2lowT, AbelEtal2012jcpH2HeCIA, RichardEtal2012jqsrtCIAs}.  
While there are newer line lists available with a greater number of lines \citep[e.g.,][]{HargreavesEtal2020apjsMethaneHITEMP}, these tests are meant to demonstrate consistency between neural-network-based and non-ML retrievals; we therefore use the setup described in \citet{HarringtonEtal2020apjsBART1}, which uses this set of line lists to compare with previous studies.
As our approach learns RT from a data set of spectra, it is not tied to any specific line lists.

To train our neural-network surrogate model, we generate 3,458,432 spectra, which are subdivided into 2,446,784 spectra (${\sim}70$\%) for training, 689,536 spectra (${\sim}20$\%) for validation, and 322,112 specra (${\sim}10$\%) for testing (for considerations about data set size, see Appendix \ref{app:datasetsize}).  
Model parameters come from the uniform distribution bound by the limits listed in Table \ref{tbl:hd189paramspace}.
Each spectrum spans 280--7100 cm$^{-1}$ at a resolution of 1.0 cm$^{-1}$ and corresponds to the planet's emitted flux in erg s$^{-1}$ cm$^{-1}$.

When processing the BART inputs/outputs for our neural network, we simplify the neural-network inputs by transforming the planet mass into the surface gravity, because this is a factor in the integration to calculate the spectrum.  
We assume a host star of radius 0.756 $R_\odot$ with a temperature of 5000 K to calculate the $T(p)$ profiles; because $\beta$ acts as a scaling factor on the related term \citep[Eq.\ 15 of][]{LineEtal2013apjRetrieval1}, it can compensate for different stellar fluxes.

We normalize the input and output data by (1) taking the logarithm of the output spectra, (2) standardizing the inputs and (log) outputs by subtracting the training mean and dividing by the training standard deviation, and (3) scaling the standardized inputs and outputs to be in the range [-1, 1].  
The neural network's input layer corresponds to the 12 inputs described above, with surface gravity replacing planetary mass.
The hidden layers consist of Conv1d(256)L(0.05) -- Dense(4096)L(0.05) -- Dense(4096)L(0.05) -- Dense(4096)L(0.05) -- Dense(4096)L(0.05). Conv1d$(n)$ indicates a 1D convolutional layer with $n$ feature maps and a kernel size of 3. L$(m)$ indicates a leaky rectified linear unit (ReLU) activation function with slope $m$ for $x < 0$.
The dense output layer has 6821 nodes, corresponding to the emitted spectrum over the defined wavenumber grid, with a ReLU activation function.
For details on our model selection process, see Appendix \ref{app:gridsearch}.

We train with a batch size of 64 using a mean-squared-error loss function, the Adam optimizer, and early stopping with a patience of 30 epochs based on the validation loss.  
We employ a cyclical learning rate that increases from $8 \times 10^{-6}$ to $5 \times 10^{-3}$ over 4 epochs, then decreases over the same window.  
After each complete cycle (8 epochs), the maximum learning rate decays by half the difference between the maximum and minimum learning rates \citep[\textit{triangular2} policy,][]{Smith2015arxivClyclicalLearningRates}.
The boundaries were chosen according to the method described in \citet{Smith2015arxivClyclicalLearningRates}, except that we consider the loss instead of accuracy (see Appendix \ref{app:gridsearch} for details).  
To evaluate the model's performance, we compute the root-mean-squared error (RMSE; comparable to the standard deviation of the differences between the predicted and true values) and the coefficient of determination ($R^2$; measures the linear correlation between the predicted and true values) between the data and the predictions, both for the full high-resolution output and the band-integrated spectra corresponding to the observations of HD 189733 b.

\subsection{Retrieval}
\label{sec:retrieval}

Following the setup of \citet{HarringtonEtal2020apjsBART1}, we perform a retrieval of the dayside atmosphere of HD 189733 b based on the measurements by the Hubble Space Telescope Near Infrared Camera MultiObject Spectrograph \citep{SwainEtal2009apjHD189733b}; Spitzer Space Telescope Infrared Spectrograph \citep[IRS][]{GrillmairEtal2008natHD189}; Spitzer InfraRed Array Camera (IRAC) channels 1 and 2 values of 0.1533 $\pm$ 0.0029\%\ and 0.1886 $\pm$ 0.0071\%\ (M. Line, priv. comm.); IRAC channel 3, IRS 16 {\micron} photometry, and Multiband Imaging Photometer for Spitzer \citep{CharbonneauEtal2008apjHD189733b}; and IRAC channel 4 \citep{AgolEtal2010apjHD189IRAC}.  
We use a K2 solar-abundance Kurucz stellar model for the host star's emission \citep{CastelliKurucz2003iausATLAS9GridModelAtmospheres}.  
Using the differential evolution Markov chain with snooker updating algorithm of \citet{Braak2008SnookerDEMC}, 2,500,000 iterations are spread across 10 parallel chains, with a burn-in of 50,000 iterations per chain.
When retrieving, we fix the semimajor axis to 0.031 au and the planetary radius and gravity at 0.1 bar to 1.138 R$_J$ and 2187.762 cm s$^{-2}$, respectively.  
The remaining neural-network input parameters are allowed to freely vary over the entire training space.

We compute the Bhattacharyya coefficient \citep{Bhattacharyya1943bcmsDivergenceProbDistn, AherneEtal1998kybBhattacharyyaMetric} to compare the similarity of 1D marginalized posteriors, where a value of 0 indicates no overlap and a value of 1 indicates identical distributions.
We choose this metric over others, such as the Kullback-–Leibler divergence, because it is both intuitive to understand and defined for all distributions, even those that do not overlap.

For this investigation, we focus on a neural network as a faster replacement for an RT code for retrieval; we therefore only compare the results of BART and the neural-network approximation.
For a discussion of these results in the context of previous retrievals of HD 189733 b's dayside atmosphere, see \citet{HarringtonEtal2020apjsBART1}.

\subsection{Software}
\label{sec:codes}

We have developed two Python packages for this investigation.  
Both are open-source software, with full documentation, under the Reproducible Research Software  License\footnote{\url{https://planets.ucf.edu/resources/reproducible-research/software-license/}}.  
We encourage users to contribute to the code via pull requests on Github.

\subsubsection{MARGE}
\label{sec:marge}
The Machine learning Algorithm for Radiative transfer of Generated Exoplanets\footnote{MARGE is available at \url{https://github.com/exosports/marge}} (MARGE, Figure \ref{fig:schematic}) (1) generates a data set based on a user-supplied function, (2) processes the generated data using a user-supplied function, and (3) trains, validates, and tests a user-specified neural-network architecture on a data set.  
The software package allows independent execution of any of the three modes, enabling a wide range of applications beyond exoplanet retrieval.

MARGE's design allows it to be applied to any deterministic model.  
For 1D data (such as spectra), MARGE's desired format is NumPy binary (.npy) files of 2D arrays, where each row corresponds to a single case.  
Each row is a data vector of the input parameters followed by the output data (e.g., spectrum).
MARGE currently includes data-generation and -processing functions for BART as well as a data-processing function for the {\tt pypsg}\footnote{\url{https://gitlab.com/frontierdevelopmentlab/astrobiology/pypsg}} Python interface \citep{SoboczenskiEtal2018arxivINARA} for the NASA Planetary Spectrum Generator \citep{VillanuevaEtal2018jqsrtPSG}.
We encourage users to contribute code via pull request to handle the processing of the inputs/outputs of other software packages.

We implement neural-network model training in Keras \citep[version 2.2.4,][]{CholletEtal2015Keras}, using a Tensorflow \citep[version 1.13.1,][]{AbadiEtal2016Tensorflow} backend.  
MARGE enables early stopping by default to prevent overfitting, and the user can halt or resume training.  
MARGE allows for cyclical learning rates for more efficient training \citep[see also Appendix \ref{app:gridsearch}]{Smith2015arxivClyclicalLearningRates}.
Users specify the model architecture details and the data location, and the software handles the data normalization, training, validation, and testing.
MARGE preprocesses the data into Tensorflow's TFRecords format for efficient handling.
Users have multiple options when preprocessing the data, which include taking the logarithm of the inputs and/or outputs, standardizing the data according to its mean and standard deviation, and/or scaling the data to be within a specified range.
The mean and standard deviation of the data set are computed using Welford's method \citep{Welford1962technomOnlineVariance} to avoid the need to load the entire data set into memory at once.
MARGE computes the RMSE and $R^2$ for predictions on the validation and test sets to evaluate model performance; these metrics can optionally be calculated over integrated filter bandpasses.
Finally, users may specify cases from the test set to plot the predicted and true spectra, with residuals (e.g., Figure \ref{fig:hd189-examples}).

\subsubsection{HOMER}
\label{sec:homer}
The Helper Of My Eternal Retrievals\footnote{HOMER is available at \url{https://github.com/exosports/homer}} (HOMER) utilizes a MARGE-trained model to infer the underlying inputs corresponding to some observed outputs (Figure \ref{fig:schematic}).
For its Bayesian framework, HOMER uses a Python wrapper for Markov chain and nested-sampling algorithms.
The user specifies data, uncertainties, observational filters, a parameter space, and a few related inputs, which are passed to the Bayesian sampler to perform the inference. 
If available, a graphics processing unit (GPU) calculates neural-network predictions, though the central processing unit (CPU) can do this at the cost of increased runtime.  
For each iteration of the Bayesian inference, the trained neural network predicts on the proposed input parameters, which are modified as necessary (descale, denormalize, divide by the stellar spectrum, unit conversions, and/or integrated over bandpasses).

HOMER produces plots of the best-fit spectrum, 1D marginalized posteriors, 2D pairwise posteriors, and parameter history traces.
The best-fit spectrum plot contains the data (with observational bandpasses indicated by uncertainties in $x$) and, if the DataSketches\footnote{\url{https://datasketches.apache.org/}} library is installed, the 1$\sigma$, 2$\sigma$, and 3$\sigma$ spectra.
We use the streaming quantiles method of \citet{KarninEtal2016focsStreamingQuantiles} as implemented in DataSketches to compute the 1-2-3$\sigma$ spectra.  
This approach avoids needing to load all of the evaluated models at once, which could exceed system memory.  

HOMER calculates the steps per effective independent sample (SPEIS) and effective sample size (ESS) as described in \citet{HarringtonEtal2020apjsBART1}.  
Markov chains make small, correlated steps; while a chain may perform 100,000 iterations, if it takes 5000 steps to materialize a completely independent sample (steps per effective independent sample, SPEIS), then there have only been 20 effective samples.  
SPEIS is calculated from the autocorrelation function of each parameter for each chain; as a conservative estimate, we use the highest SPEIS value when calculating the ESS of the Bayesian inference to ensure we do not underestimate credible region uncertainties. 
By rearranging Equation 1 of \citet{HarringtonEtal2020apjsBART1}, an uncertainty $s_{\hat{C}}$ can be calculated on a given credible region $\hat{C}$ based on the ESS:
\begin{equation}
    s_{\hat{C}} \approx \sqrt{\frac{\hat{C}(1 - \hat{C})}{ESS}}
\end{equation}
For example, if the ESS is 20, then the determined 68.27\% credible region is actually the 68.27 $\pm$ 10\% credible region; running the inference for more iterations would increase the ESS and accordingly decrease the uncertainty on that credible region.

For easy comparison with other retrieval results, HOMER can overplot the 1D and 2D posteriors for multiple retrievals (e.g., Figure \ref{fig:homer-hd189}) and compute the Bhattacharyya coefficients between the 1D posteriors.

\section{Results \& Discussion}
\label{sec:results}

The normalized RMSE, normalized $R^2$, and denormalized $R^2$ metrics for the MARGE-trained model on the test set for the high-resolution and band-integrated spectra are detailed in Tables \ref{tbl:marge-eval} and \ref{tbl:marge-eval-integ}, respectively.  
The normalized RMSE ${\ll}1$ and $R^2{\sim}1$ indicate an accurate model for RT over the parameter space.  
Rather than waiting for early stopping to engage, we manually stopped training at 130 epochs because there was an insignificant improvement in the loss for dozens of epochs.
For considerations on how this affects model performance, see Appendix \ref{app:datasetsize}.

Figure \ref{fig:hd189-examples} shows example comparisons between the spectra predicted by MARGE and true spectra calculated by BART.  
While residuals tend to be around a few percent, they generally fluctuate around 0; when band-integrated over the observational filters, these errors usually cancel, as shown by the lower normalized RMSE and higher denormalized $R^2$ metrics (Tables \ref{tbl:marge-eval} and \ref{tbl:marge-eval-integ}).
We observe that in some cases, there are regions where the spectrum is consistently over- or underestimated by a few percent (e.g., the top-left panel of Figure \ref{fig:hd189-examples} around 4250 cm$^{-1}$), thereby introducing error in the band-integrated value.
However, the small deviations appear to have only a minor effect on this retrieval's result; see Section \ref{sec:limitations} for considerations when retrieving at high spectral resolutions or in cases where a traditional retrieval result is not available for comparison.

\begin{table}[tb]
\caption{Model Evaluation: High-resolution Spectra}\vskip -.25in
\label{tbl:marge-eval}
\begin{center}
\begin{tabular}{l c c c c }
 \toprule
Metric & Min. & Median & Mean & Max. \\
\hline
Norm. RMSE    & 0.00153 & 0.00224 & 0.00247 & 0.01040 \\
Norm. $R^2$   & 0.99999 & 1.00000 & 1.00000 & 1.00000 \\
Denorm. $R^2$ & 0.99885 & 0.99993 & 0.99990 & 0.99997 \\
\hline
\multicolumn{5}{p{0.95\linewidth}}{\textbf{Notes.}  RMSE and $R^2$ are calculated for each of the 6821 outputs corresponding to the wavenumber grid of 280--7100 cm$^{-1}$ with a resolution of 1.0 cm$^{-1}$.  For conciseness, we present statistics about these values.  The $R^2$ values are slightly less than 1, but they round to 1 at the reported precision.}
\end{tabular}
\end{center}
\end{table}

\begin{table}[tb]
\caption{Model Evaluation: Band-integrated Spectra}\vskip -.25in
\label{tbl:marge-eval-integ}
\begin{center}
\begin{tabular}{l c c c c }
 \toprule
Metric & Min. & Median & Mean & Max. \\
\hline
Norm. RMSE    & 0.00123 & 0.00147 & 0.00148 & 0.00183 \\
Norm. $R^2$   & 1.00000 & 1.00000 & 1.00000 & 1.00000 \\
Denorm. $R^2$ & 0.99995 & 0.99997 & 0.99997 & 0.99998 \\
\hline
\multicolumn{5}{p{0.95\linewidth}}{\textbf{Notes.}  Same as Table \ref{tbl:marge-eval}, except integrated over the 66 bandpasses corresponding to the referenced observations of HD 189733 b.}
\end{tabular}
\end{center}
\end{table}

\begin{figure*}[htb]
\centering
\includegraphics[width=0.49\textwidth, clip]{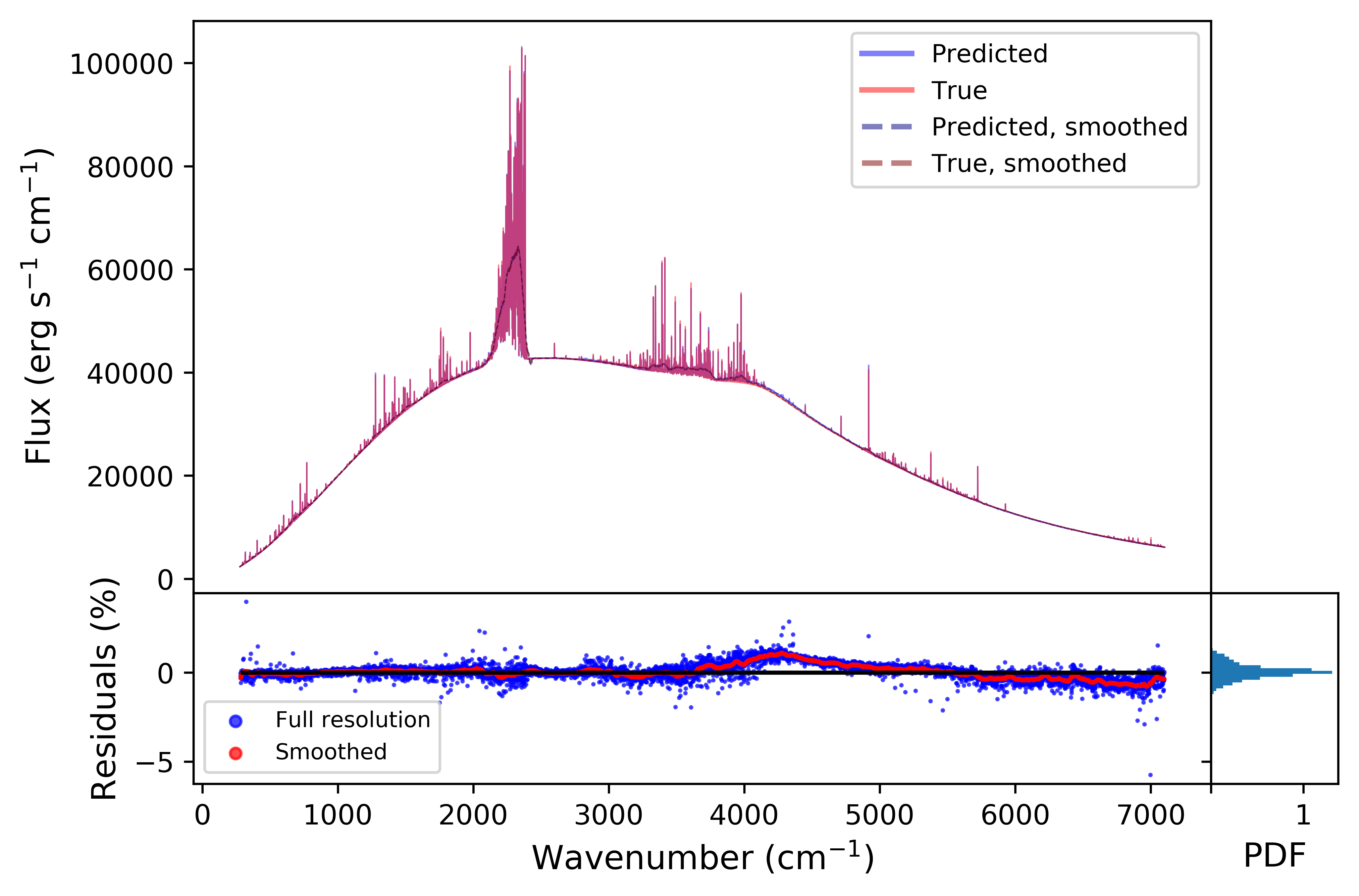}\hfill
\includegraphics[width=0.49\textwidth, clip]{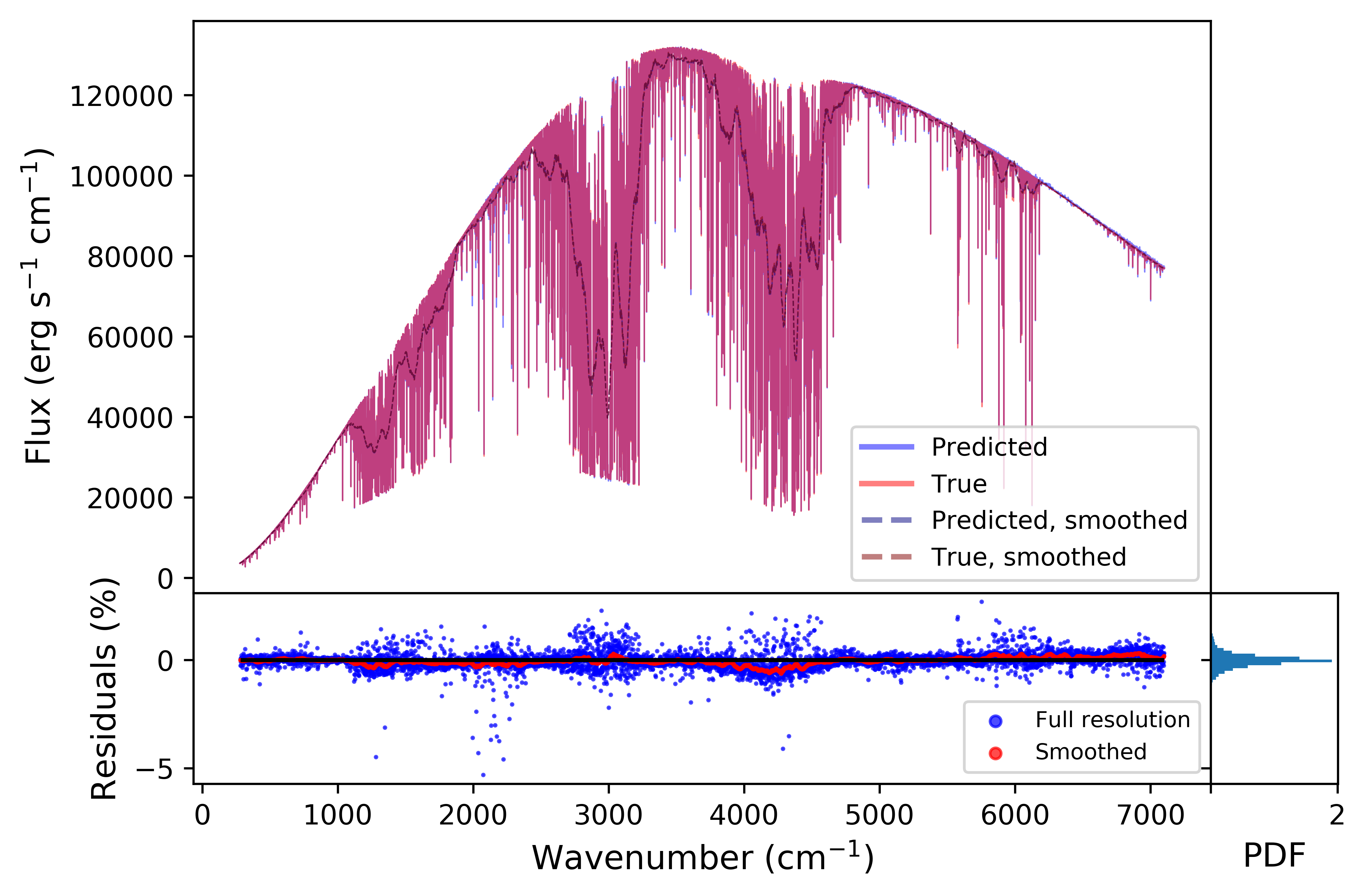}\\
\includegraphics[width=0.49\textwidth, clip]{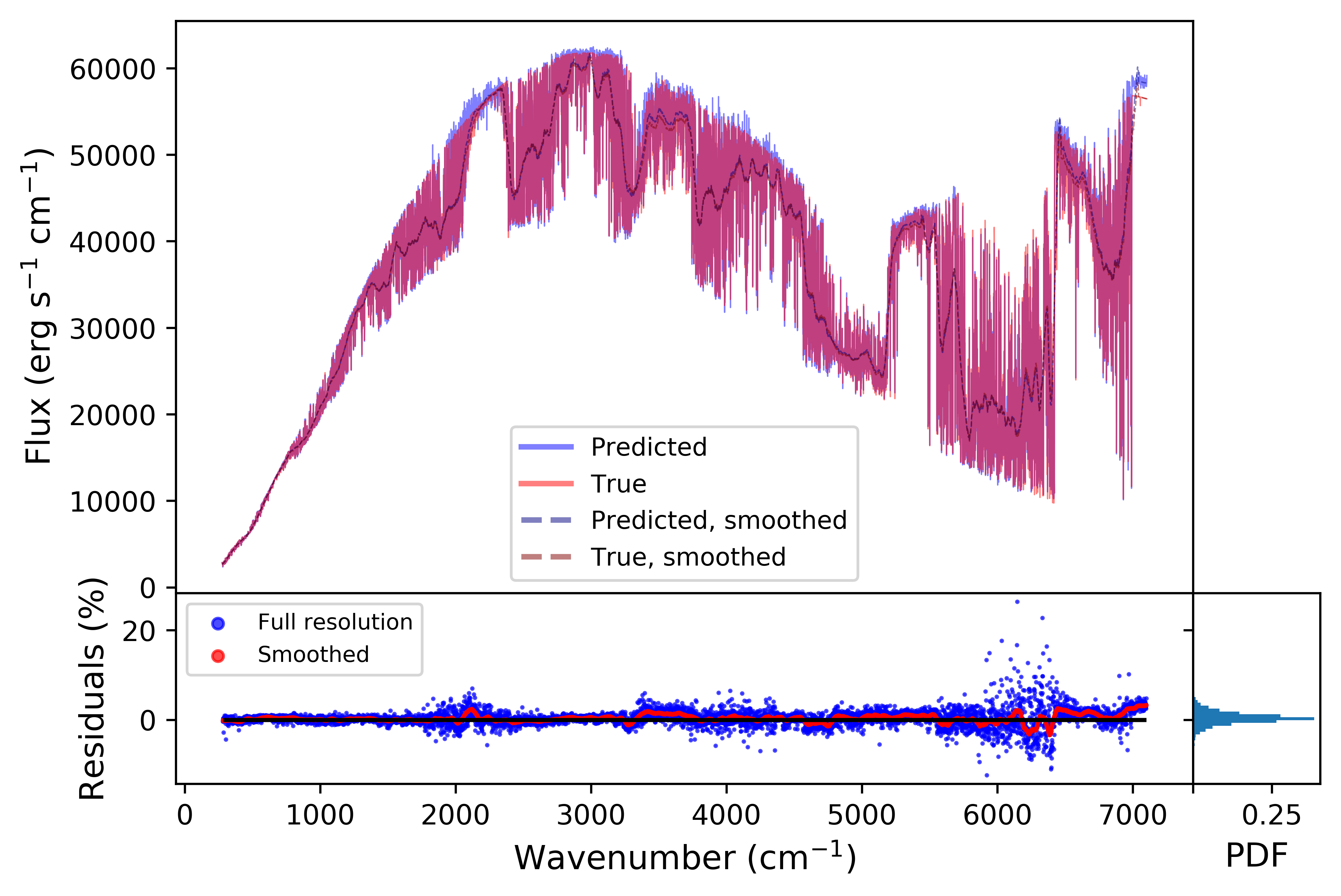}\hfill
\includegraphics[width=0.49\textwidth, clip]{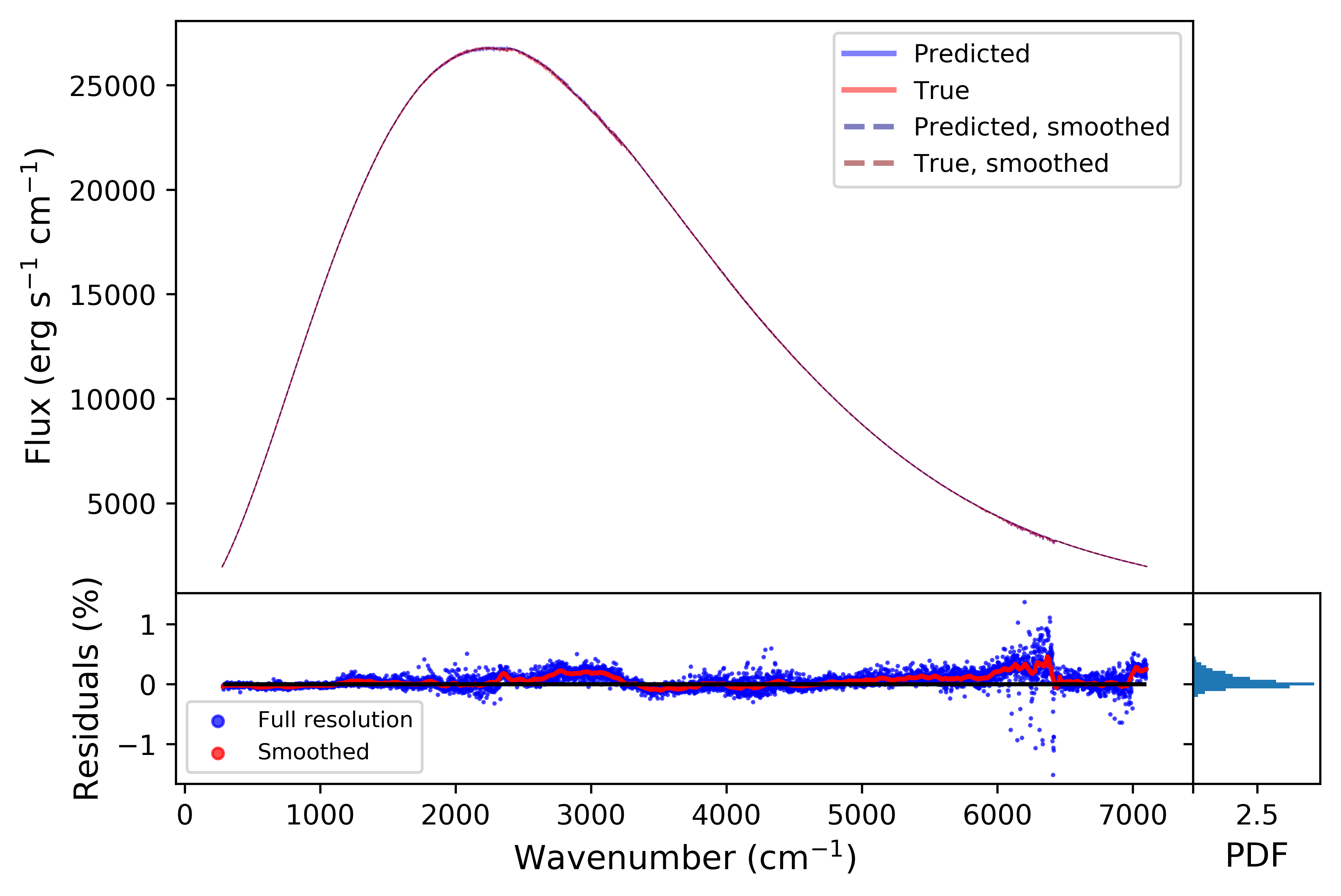}\\
\caption{Four comparisons of planetary emission spectra predicted by MARGE and calculated by BART.  
The smoothed curves use a Savitzky-Golay filter with a third-order polynomial across a window of 101 elements (100 cm$^{-1}$).  
The purple color arises due to a detailed match between the red and blue spectra at high resolution.  
For the residuals, a black line is plotted at 0 to show regions where the neural network consistently over- or underpredicts the spectrum.  
A histogram of the high-resolution residuals appears to the right of the residual scatter plot, where the x-axis shows the probability density function (PDF) for the range of residual percentages.  
\textbf{Top left:} Case with $T(p)$ profile that increases in temperature with altitude, with H$_2$O and CO$_2$ emission lines.  
\textbf{Top right:} Case with $T(p)$ profile that decreases in temperature with altitude, with absorption primarily due to CH$_4$ and H$_2$O.  
\textbf{Bottom left:} Cases with $T(p)$ profile that has an inversion around 0.1 bar, with CH$_4$, CO, and CO$_2$ absorption and emission features.  
\textbf{Bottom right:} Case with $T(p)$ profile that is nearly isothermal at the pressures with sensitivity.
\label{fig:hd189-examples}}
\end{figure*}

\begin{figure*}[htb]
\centering
\includegraphics[width=0.54\textwidth, clip]{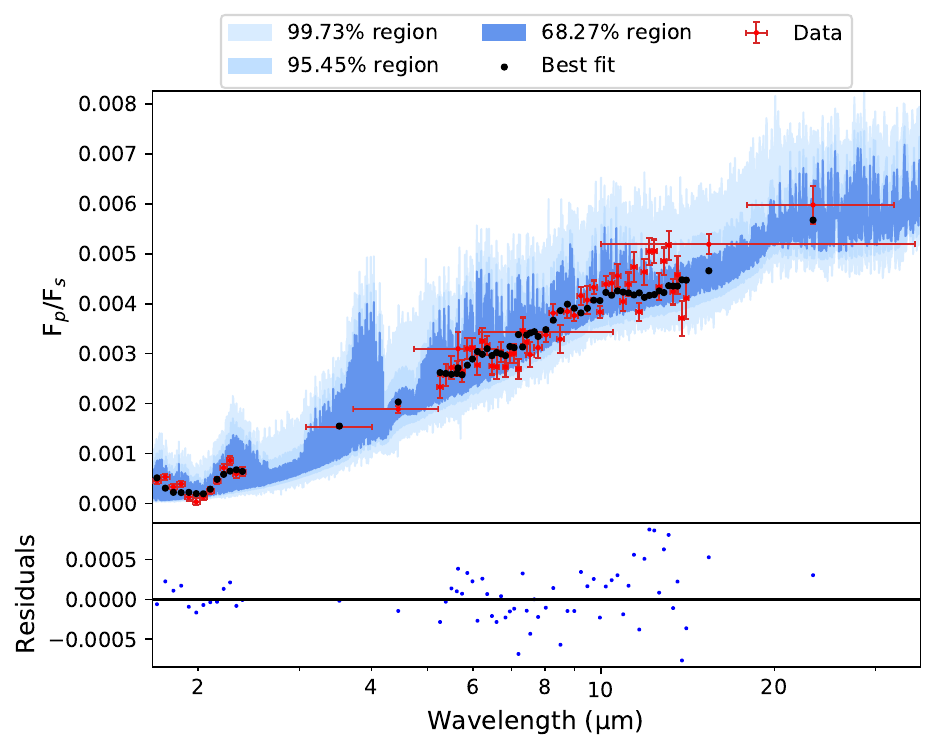}\hfill
\includegraphics[width=0.45\textwidth, clip]{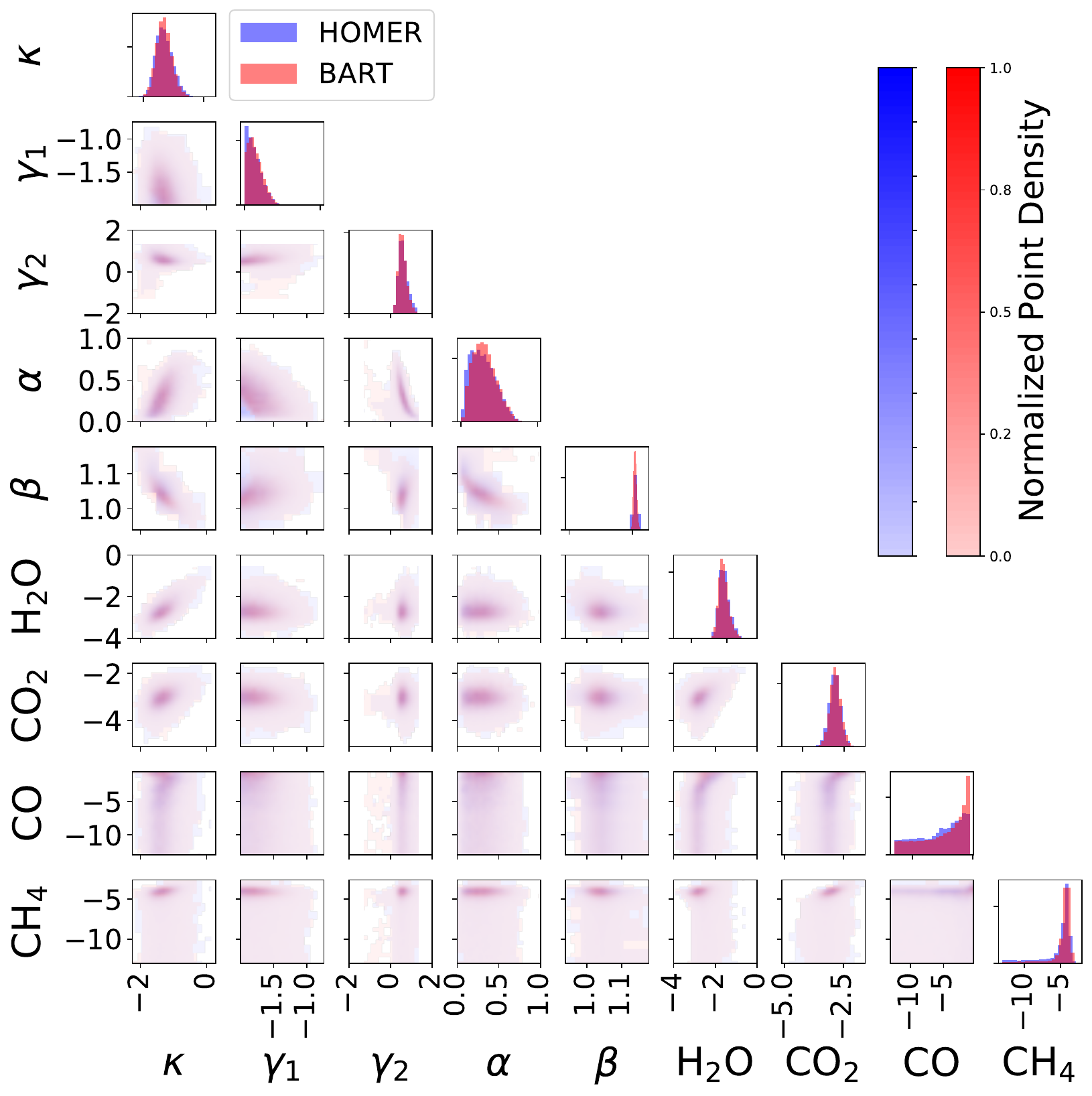}\\
\includegraphics[width=0.515\textwidth, clip]{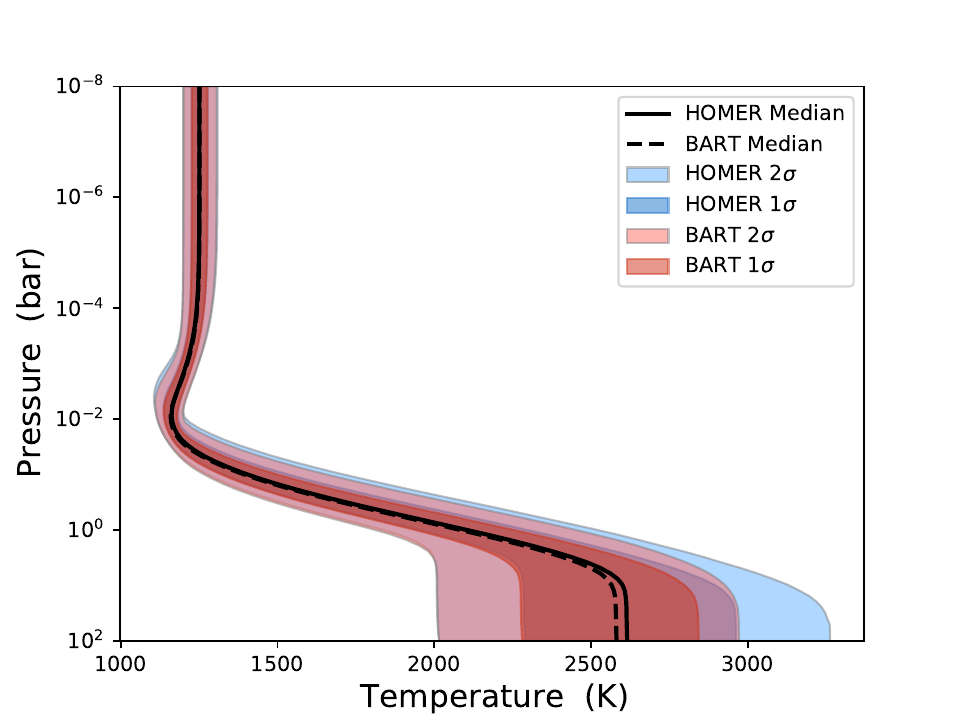}\hfill
\includegraphics[width=0.47\textwidth, clip]{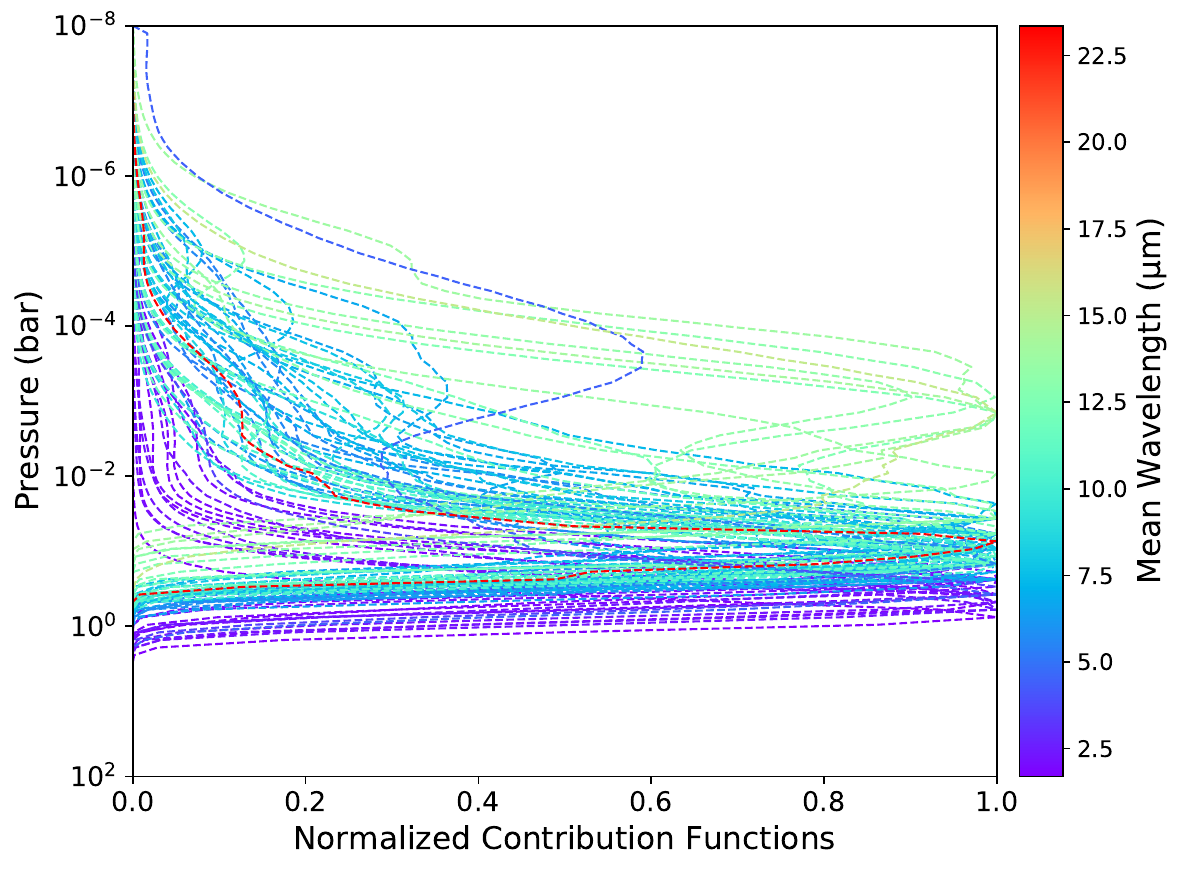}
\caption{Comparisons between HOMER and BART posteriors.  Top left: Best-fit spectrum of HOMER, with 1, 2, and 3$\sigma$ regions. Top right: Normalized probability density functions of the 2D marginalized pairwise posteriors retrieved for HD 189733 b, with the 1D marginalized posteriors along the diagonal.  The purple color arises from the close match between HOMER and BART.  Bottom left: Posterior median, 1$\sigma$ from the median, and 2$\sigma$ from the median $T(p)$ profile.  In the regions with sensitivity, HOMER closely matches BART, with a slightly greater uncertainty.  Bottom right: Normalized contribution functions, which show the pressure range each filter probes, for the best-fit BART model.}
\label{fig:homer-hd189}
\end{figure*}

When applying HOMER to the emission spectrum of HD 189733 b, the results are consistent with BART.  
The retrieved $T(p)$ profiles (bottom-left panel Figure \ref{fig:homer-hd189}) agree in the regions probed by the observations (\textless1 bar, bottom-right panel Figure \ref{fig:homer-hd189}) and only begin to deviate deeper in the atmosphere, where little to no signal is measured according to the contribution functions.  
By nature, HOMER cannot calculate contribution functions, as the MARGE model does not solve RT.  
While they could be included for each case in the training set, this would require significantly more compute resources.
Computing the contribution functions for the single best-fit case using the RT code that trained MARGE more efficiently uses compute resources.

Table \ref{tbl:credreg-hd189} compares HOMER's retrieved 68.27\% (``1$\sigma$''), 95.45\% (``2$\sigma$''), and 99.73\% (``3$\sigma$'') credible regions with BART's retrieved credible regions.
All regions closely agree, with differences attributable to a combination of uncertainty from a finite ESS and the neural network's imperfect nature (Figure \ref{fig:homer-hd189}, top-right panel).
For CO, both BART and HOMER favor large abundances, though BART finds a greater probability for log abundances $\geq -2$ (Figure \ref{fig:homer-hd189}, top-right panel).  
Despite this, the credible regions agree (Table \ref{tbl:credreg-hd189}). 
Similarly, HOMER favors lower values for $\gamma_1$ and $\alpha$, though the resulting thermal profiles agree (Figure \ref{fig:homer-hd189}, bottom-left panel).  

Table \ref{tbl:ess} compares the SPEIS, ESS values, and associated uncertainties in the 1$\sigma$, 2$\sigma$, and 3$\sigma$ credible regions for HOMER and BART.  
HOMER yields an SPEIS that is less than BART's, attributable to the conservative estimate of SPEIS as being the greatest among all chains and parameters.  
The highest SPEIS values fluctuate between runs, though the median SPEIS remains relatively constant.
HOMER's median SPEIS of 627 and BART'S 615 better reflect the close agreement between the two retrievals.
The Bhattacharyya coefficients between the 1D marginalized posteriors of HOMER and BART indicate agreement in the range 0.9843--0.9972, with a mean of 0.9925 (Table \ref{tbl:bhatchar}).  

\begin{table*}[thb]
\caption{Retrieved Credible Regions}\vskip -.1in
\label{tbl:credreg-hd189}
\begin{center}
\begin{tabular}{ l l c c c }
 \toprule
Parameter        & Code & 68.27\% & 95.45\% & 99.73\% \\
\hline
log $\kappa$     & HOMER & [-1.63, -1.06] & [-1.84, -0.71] & [-2.07, -0.33] \\
                 & BART  & [-1.58, -1.09] & [-1.81, -0.79] & [-1.99, -0.46] \\
log ${\gamma}_1$ & HOMER & [-1.98, -1.65] & [-1.99, -1.34] & [-1.99, -1.06] \\
                 & BART  & [-1.98, -1.62] & [-2.00, -1.33] & [-2.00, -1.07] \\
log ${\gamma}_2$ & HOMER & [0.34, 0.77] & [0.21, 1.10] & [ 0.11, 1.29] \\
                 & BART  & [0.35, 0.73] & [0.21, 1.02] & [-0.07, 1.30] \\
$\alpha$         & HOMER & [0.07, 0.39] & [0.03, 0.60] & [0.01, 0.74] \\
                 & BART  & [0.11, 0.42] & [0.06, 0.60] & [0.02, 0.74] \\
$\beta$          & HOMER & [1.01, 1.07] & [0.99, 1.12] & [0.96, 1.15] \\
                 & BART  & [1.01, 1.06] & [0.99, 1.10] & [0.97, 1.15] \\
log H$_2$O       & HOMER & [-3.11, -2.37] & [-3.37, -1.82] & [-3.70, -1.27] \\
                 & BART  & [-3.12, -2.44] & [-3.37, -1.92] & [-3.63, -1.41] \\
log CO$_2$       & HOMER & [-3.39, -2.73] & [-3.78, -2.36] & [-4.26, -2.01] \\
                 & BART  & [-3.33, -2.71] & [-3.66, -2.32] & [-4.05, -2.03] \\
log CO           & HOMER & [-6.89, -0.51] & [-12.02, -0.51] & [-12.90, -0.51] \\
                 & BART  & [-6.60, -0.50] & [-12.55, -0.50] & [-12.90, -0.50] \\
log CH$_4$       & HOMER & [-5.16, -3.53] & [-10.25, -3.20 & [-12.95, -3.12] \\
                 & BART  & [-4.71, -3.67] & [-10.53, -3.14] & [-12.73, -3.07] \\
\hline
\end{tabular}
\end{center}
\end{table*}

\begin{table*}[tb]
\caption{Credible Region Accuracy}\vskip -.1in
\label{tbl:ess}
\begin{center}
\begin{tabular}{ l l l l l l}
 \toprule
Code & SPEIS & ESS & 1$\sigma$ Uncertainty & 2$\sigma$ Uncertainty & 3$\sigma$ Uncertainty \\
\hline
HOMER & 1668 & 1199 & 1.34\% & 0.60\% & 0.15\% \\
BART  & 2084 &  959 & 1.50\% & 0.67\% & 0.17\% \\
\hline
\end{tabular}
\end{center}
\end{table*}

\begin{table}[tb]
\caption{Bhattacharyya Coefficients}\vskip -.1in
\label{tbl:bhatchar}
\begin{center}
\begin{tabular}{ l l}
 \toprule
Parameter & Value \\
\hline
$\kappa$   & 0.9948 \\
$\gamma_1$ & 0.9972 \\
$\gamma_2$ & 0.9950 \\
$\alpha$   & 0.9909 \\
$\beta$    & 0.9879 \\
H$_2$O     & 0.9968 \\
CO$_2$     & 0.9968 \\
CO         & 0.9888 \\
CH$_4$     & 0.9843 \\
Mean       & 0.9925 \\
\hline
\end{tabular}
\end{center}
\end{table}

\subsection{Limitations}
\label{sec:limitations}

HOMER's accuracy is, by nature, bound by the accuracy of the neural-network model.
Model inaccuracies may slightly bias the results, as seen in the minor differences between the posteriors of HOMER and  BART.
In our application, this discrepancy does not significantly affect the scientific conclusions at the spectral resolution of these observations for the current neural-network accuracy.  
However, this does not necessarily hold for all cases.  
It is possible that at higher resolutions this neural network's minor inaccuracies can drive the Bayesian sampler to radically different results. 
While in theory MARGE works for any spectral resolution, users will need to carefully select the model architecture to ensure that it can accurately model the spectra over the desired phase space.  
In situations lacking a physics-based retrieval to compare with, we advise testing to ensure that forward models are reasonably accurate over the retrieval's phase space, as some regions may not be sufficiently sampled for accurate predictions.  

\subsection{Compute Cost}
\label{sec:compcost}

The performance differences between HOMER and BART highlight HOMER's computational benefits.  
For a single Markov chain iteration, BART requires around 1.8 seconds per parallel chain on an AMD EPYC 7402P CPU, and multiple chains parallelize linearly across CPUs.  
By comparison, a single iteration with HOMER on the same CPU --- which includes preprocessing (e.g., input normalization), prediction, post-processing (e.g., output denormalization, scaling according to the stellar spectrum), and band-integration --- requires just $\sim0.2$ s for any number of chains fewer than 32.
For a single chain, this is thus a $\sim9\times$ speedup.  
In our setup, we considered 10 parallel chains, translating to a $\sim90\times$ speedup for the function evaluated at each step of the Markov chain.  
Using an NVIDIA Titan Xp for predictions, the model evaluations at each Markov chain step require 0.01--0.02 seconds, a 10--20$\times$ speedup over predictions with the aforementioned CPU and, when using 10 parallel chains, a 900--1800$\times$ speedup over the same function evaluation in BART.
We note that if BART were capable of utilizing a GPU, this speedup factor would be much less.  
Further investigation is necessary to determine whether HOMER offers speed improvements over a GPU-accelerated RT code.  
Nevertheless, the CPU results emphasize the speed improvements of our approach; the significant reduction in compute time enables retrievals to be executed on an average laptop.
We note that the memory footprints for both approaches were comparable, though the parameters of each approach can strongly affect the required memory.

Here, the upfront compute cost to generate a data set and train a MARGE model is greater than the time to execute a single BART retrieval.  
In our example, we generated around 1.5--2$\times$ the number of spectra typically computed during a BART retrieval with small credible region uncertainties, plus a few dozen hours to train the neural network.  
However, additional retrievals within the trained parameter space execute in around 30 minutes on our GPU (less when neglecting to compute spectral quantiles).
Thus, when carrying out even two retrievals within some shared phase space, the compute cost of MARGE+HOMER is less than two classical retrievals.
In certain circumstances, such as where the radius and mass of the planet do not need to be varied (e.g., retrievals on different data sets of the same exoplanet), the number of spectra required to approximate the phase space accurately would be less than in our example, which may lead to MARGE+HOMER requiring less compute time than a single BART retrieval.
Beyond the scope of retrieval, this approach could also provide a benefit to situations where it is advantageous to trade one set of computing resources for another. 
For example, spaceflight missions may be limited by thermal, power, and/or on-board computational resources; it may be advantageous to increase the total compute time, if it can decrease the power, thermal, and/or on-board computing required for the calculation.

Another benefit of our approach is that the compute-cost scaling is less than linear: increasing from 10 to 256 chains results in just a $\sim12.5\times$increase in compute cost per iteration when using a GPU, compared to 25.6$\times$ as much for BART. 
Additional chains enable faster exploration of the parameter space, and, if executed for the same number of iterations per chain, increases the ESS, which reduces the uncertainty in the bounds of credible regions \citep{HarringtonEtal2020apjsBART1}.  
Thus, the combination of MARGE and HOMER saves valuable compute resources when performing retrievals and reduces total runtime when performing multiple retrievals.

\section{Conclusions}
\label{sec:conclusions}

This paper presents a novel technique for ML retrieval that uses a neural-network model of RT within a Bayesian framework to reduce the runtime of a retrieval.
Our open-source codes, MARGE and HOMER, provide the community with an easy-to-use implementation of this approach, and they are readily applicable to any forward model and its inversion --- not strictly BART or even RT. 
They are available on Github with full user documentation. 

Our method enables fast retrievals that are consistent with algorithms that solve the RT equation.  
The approach circumvents limitations of current ML retrieval models by using an RT surrogate in place of the RT code found in classical retrieval algorithms, thereby preserving the accuracy of the Bayesian inference.  
Like BART, MARGE and HOMER work at both the low resolutions of Spitzer and the high spectral resolutions of advanced ground-based spectrographs.

On our hardware, HOMER reduces the runtime of each MCMC iteration by $\sim9\times$ per parallel chain using a CPU and 90--180$\times$ per chain using a GPU, compared to BART.
For the case of HD 189733 b, the Bhattacharyya coefficients of the 1D marginalized posteriors of BART and HOMER are {\textgreater}0.984, indicating a close match.
This reduction in compute time enables using more realistic (and computationally expensive) RT models, such as those including scattering and condensates.
Additionally, 3D retrievals with $\sim$200 cells could be completed in a matter of days.  

Our approach is particularly well suited to planning studies for future observations, telescopes, and instruments, like the James Webb Space Telescope and the Large UltraViolet Optical InfraRed Surveyor \citep[e.g.,][]{RochettoEtal2016apjJWSTRetrievals, FengEtal2018ajFutureTelescopesRetrievals}.
Using a single MARGE model trained over the desired parameter space, HOMER can perform dozens to hundreds of retrievals in the time it takes to run a single retrieval with an RT solver.

More generally, our technique and tools can be applied to problems beyond the scope of this investigation.  
MARGE provides a generalized method to train a neural network to model any deterministic process, while HOMER uses a MARGE-trained model to infer the inverse process.
MARGE models could be trained for cloud/haze formation or photochemistry within general circulation models, for example.  
MARGE and HOMER could also be used to map gravitationally lensed galaxies \citep[e.g.,][]{PerreaultLevasseurEtal2017apjlGravitationalLensing}. 

With the plethora of ML retrieval algorithms that have emerged in recent years, standard data sets should be created and used for benchmarking.  
Ideally, such a data set would cover a wide range of wavelengths at high resolution and include all available opacity sources, scattering, clouds/hazes, and, in the case of terrestrial planets, surface properties. 
This would allow easy comparisons among current and future ML retrieval codes.

The Reproducible Research Compendium for this work is available for download\footnote{Available at \url{https://exoplanetarchive.ipac.caltech.edu/docs/marge-homer.html}}.  
It includes all of the code, configuration files, data, and plots used in support of this work.

\acknowledgements

We gratefully acknowledge Jon Malkin, Lee Rhodes, and Edo Liberty for valuable contributions to the streaming quantiles method used in this work.  
We thank James Mang and Nicholas Susemiehl for useful feedback on the software developed for this work.  
We thank Michael Lund and the NASA Exoplanet Archive for preparing and hosting the online RRC.
We thank Jennifer Adams for helpful discussions on radiative transfer emulation in Earth science.  
We also thank contributors to the Datasketches library, NumPy, SciPy, Matplotlib, Tensorflow, Keras, the Python Programming Language, the free and open-source community, and the NASA Astrophysics Data System for software and services. 
We gratefully acknowledge the support of NVIDIA Corporation with the donation of the Titan Xp GPU used for this research.
This research was supported by the NASA Fellowship Activity under NASA Grant 80NSSC20K0682 and NASA Exoplanets Research Program grant NNX17AB62G.
We thank FDL (http://www.frontierdevelopmentlab.org/) and SETI (https://www.seti.org) for making this collaboration possible.

\appendix
\counterwithin{figure}{section}
\counterwithin{table}{section}

\section{Determining Model Architecture}
\label{app:gridsearch}

To select an ideal neural-network architecture, a grid search must be carried out.  
This includes varying the types of hidden layers, number of hidden layers, number of nodes per layer, activation functions for each hidden layer, the parameter(s) for each activation function, and the learning rate.

We carried out a grid search by training each model on a subset of the total data set (171,456 training, 66,432 validation) for 20 epochs using a batch size of 64.  
We considered 3--5 dense and convolutional+pooling hidden layers; 64--4096 nodes; rectified linear unit (ReLU), leaky ReLU, exponential linear unit (ELU), hyperbolic tangent (tanh), and sigmoid activation functions.  
The convolutional layers use a kernel size of 3, and pooling layers use a size of 2.
We consider four learning rate policies: (1) a cyclical rate ranging from 8 $\times 10^{-6}$ -- 5 $\times 10^{-3}$ where the maximum is reduced by half of the difference with the minimum every 8 epochs, (2) as before but ranging from $10^{-5}$ -- $10^{-3}$, (3) a constant learning rate of $10^{-5}$, and (4) a constant $10^{-3}$.
Policies 3 and 4 are only considered for models that do not include tanh or sigmoid activations.

Table \ref{tbl:gridsearch} presents the minimum validation loss for each architecture considered.
There is some randomness to the minimum validation loss due to the shuffling of the training data, so models with comparable minimum validation losses can be considered equivalent in performance.
We chose to perform a more exhaustive grid search than is typical to emphasize certain points that can guide future investigations.

In general, we find that models with 4+ hidden layers with ReLU, leaky ReLU, and ELU activations achieve the lowest validation loss for this problem.
The best-performing models all have a 1D convolutional layer as the first hidden layer, while the worst-performing models use tanh or sigmoid activations.  
Additional layers generally lead to reductions in the loss. 
Cases where this does not occur can be attributed to the learning rate policy (e.g., models 25--27, LR1 vs. LR2), highlighting the importance of properly selecting the policy (described below).  
Minor variations (e.g., models 28--30 LR1) can be attributed to training randomness.
ReLU and leaky ReLU activations tend to have similar performance; leaky ReLU with a small parameter tends to perform equivalently or better than ReLU (e.g., models 32 and 33).
While these results point to deep architectures as optimal configurations for this application of ML RT, tests varying the spectral resolution, wavelength range, etc. are necessary to definitively confirm if such variations change the optimal architecture(s).
A future investigation should consider this in more detail.

Based on this grid search, we selected model 40.  
While a similar architecture with ELU activations performed equivalently (model 37), it took longer to train per epoch.  
Additionally, we found that the retrieval accuracy did not significantly change below some threshold validation loss (see Appendix \ref{app:datasetsize}), so training time is a more important consideration than minor differences in minimum validation loss.

Our results show that, when the learning rate range is properly chosen, cyclical learning rates outperform constant learning rates, confirming the findings of \citet{Smith2015arxivClyclicalLearningRates} for this particular problem.
Select models do not follow this trend (e.g., models 31, 35, 38), which is likely attributable to the small number of epochs considered in this grid search.

\begin{figure}[!h]
\centering
\includegraphics[trim={0 0 0 0.64cm}, width=0.49\linewidth, clip]{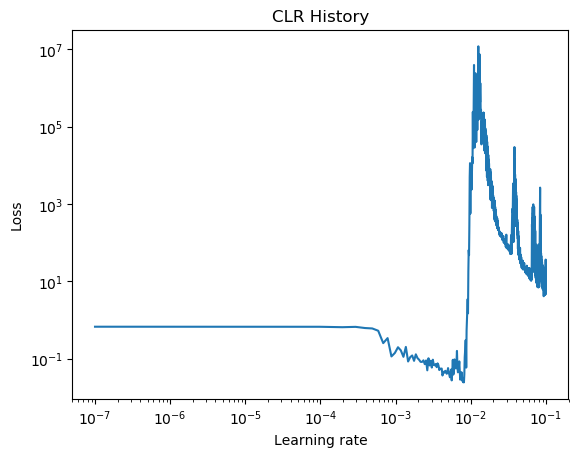}
\caption{Example of a range test.  The learning rate begins at a value too small to make noticeable changes to the weights of the model.  At a learning rate of $\sim 4 \times 10^{-4}$, the loss begins to decrease, indicating that the model has begun learning.  However, at a learning rate $\sim  5 \times 10^{-3}$, the loss begins to behave erratically, and it becomes very large at a learning rate of $10^{-2}$.  From this, a learning rate policy varying between $6 \times 10^{-4}$ and $4 \times 10^{-3}$ would likely perform well for this architecture.}
\label{fig:rangetest}
\end{figure}

In Section \ref{sec:methods}, we note that we make the learning rate policy selection as described in \citet{Smith2015arxivClyclicalLearningRates}, except based on the loss instead of the accuracy.  
Our selection process is to perform a `range test' by training the model over a few epochs using a learning rate policy that constantly increases from a very small rate (e.g., $10^{-7}$) to a large rate (e.g., $10^{-1}$).  
Looking at a plot of loss \textit{vs.} learning rate (e.g., Figure \ref{fig:rangetest}), the learning rate range can be deduced based on when the loss begins decreasing (minimum learning rate) and when the loss begins increasing (maximum learning rate).
In practice, we find more efficient training using a range that is slightly interior to the extrema determined via the plot.  
This is analogous to the method described in \citet{Smith2015arxivClyclicalLearningRates}, except that it is more straightforward to determine the learning rate boundaries.

\begin{longtable}{r  l  r r r r}
\caption{Model Grid Search, 20 Epochs\label{tbl:gridsearch}}\\
 \toprule
\# & Hidden Layers & \multicolumn{4}{c}{Min. Val. Loss ($\times 10^5$)}\\
   &               & LR1$^*$ & LR2$^{\dagger}$ & LR3$^{\ddagger}$ & LR4$^{\mathsection}$ \\
\hline\endhead
\rownumber & D(512)$^a$R$^b$--D(512)R--D(512)R      &   17.4  & 43.8 & 491  & 19.2\\
\rownumber & D(1024)R--D(1024)R--D(1024)R           &    9.61 & 24.6 & 291  & 12.3\\
\rownumber & D(2048)R--D(2048)R--D(2048)R           &    7.60 & 13.3 & 179  & 8.28\\
\rownumber & D(4096)R--D(4096)R--D(4096)R           &    8.56 & 7.17 & 106  & 6.54\\
\rownumber & D(512)R--D(512)R--D(512)R--D(512)R     &   13.0  & 31.3 & 382  & 16.2\\
\rownumber & D(512)S$^c$--D(512)S--D(512)S--D(512)S &   68.0  & 951  & ---  & ---\\
\rownumber & D(512)T$^d$--D(512)T--D(512)T--D(512)T &  487    & 47.0 & ---  & ---\\
\rownumber & D(512)S--D(1024)S--D(2048)S--D(4096)S  &   48.5  & 932  & ---  & ---\\
\rownumber & D(512)T--D(1024)T--D(2048)T--D(4096)T  & 1390    & 68.8 & ---  & ---\\
\rownumber & D(1024)R--D(1024)R--D(1024)R--D(1024)R &    8.43 & 16.8 & 238  & 10.1\\
\rownumber & D(2048)R--D(2048)R--D(2048)R--D(2048)R &    7.46 & 9.13 & 125  & 8.01\\
\rownumber & D(4096)R--D(4096)R--D(4096)R--D(4096)R &    8.23 & 5.05 & 62.5 & 6.14\\
\rownumber & D(4096)S--D(4096)S--D(4096)S--D(4096)S & 1350    & 197  & ---  & ---\\
\rownumber & D(4096)T--D(4096)T--D(4096)T--D(4096)T & 1740    & 46.0 & ---  & ---\\
\rownumber & D(4096)E(0.05)$^e$--D(4096)E(0.05)--D(4096)E(0.05)--D(4096)E(0.05)  & 220    & 5.10 & 64.6 & 5.55\\
\rownumber & D(4096)E(0.1)--D(4096)E(0.1)--D(4096)E(0.1)--D(4096)E(0.1)          & 227    & 5.03 & 70.3 & 5.93\\
\rownumber & D(4096)E(0.15)--D(4096)E(0.15)--D(4096)E(0.15)--D(4096)E(0.15)      & 206    & 5.13 & 74.5 & 7.13\\
\rownumber & D(4096)E(0.2)--D(4096)E(0.2)--D(4096)E(0.2)--D(4096)E(0.2)          & 238    & 5.39 & 81.7 & 6.18\\
\rownumber & D(4096)L(0.05)$^f$--D(4096)L(0.05)--D(4096)L(0.05)--D(4096)R        &   7.32 & 5.21 & 67.2 & 7.33\\
\rownumber & D(4096)L(0.05)--D(4096)L(0.05)--D(4096)L(0.05)--D(4096)L(0.05)      & 232    & 4.96 & 65.5 & 6.82\\
\rownumber & D(4096)L(0.1)--D(4096)L(0.1)--D(4096)L(0.1)--D(4096)L(0.1)          & 270    & 5.09 & 72.2 & 6.13\\
\rownumber & C(64)$^g$E(0.05)--M$^h$(2)--D(4096)E(0.05)--D(4096)E(0.05)                          & 5.89  & 9.69 & 123  & 6.34\\
\rownumber & C(64)E(0.05)--M(2)--D(4096)E(0.05)--D(4096)E(0.05)--D(4096)E(0.05)                  & 4.71  & 5.67 & 70.6 & 4.43\\
\rownumber & C(64)E(0.05)--M(2)--D(4096)E(0.05)--D(4096)E(0.05)--D(4096)E(0.05)--D(4096)E(0.05)  & 4.76  & 4.58 & 51.7 & 5.79\\
\rownumber & C(64)L(0.05)--M(2)--D(4096)L(0.05)--D(4096)L(0.05)                                  & 6.06  & 9.61 & 118  & 6.41\\
\rownumber & C(64)L(0.05)--M(2)--D(4096)L(0.05)--D(4096)L(0.05)--D(4096)L(0.05)                  & 4.55  & 5.56 & 71.5 & 4.71\\
\rownumber & C(64)L(0.05)--M(2)--D(4096)L(0.05)--D(4096)L(0.05)--D(4096)L(0.05)--D(4096)L(0.05)  & 149   & 4.57 & 49.3 & 4.24\\
\rownumber & C(128)E(0.05)--M(2)--D(4096)E(0.05)--D(4096)E(0.05)                                 & 5.94  & 8.49 & 105  & 5.04\\
\rownumber & C(128)E(0.05)--M(2)--D(4096)E(0.05)--D(4096)E(0.05)--D(4096)E(0.05)                 & 4.77  & 5.19 & 63.5 & 4.40\\
\rownumber & C(128)E(0.05)--M(2)--D(4096)E(0.05)--D(4096)E(0.05)--D(4096)E(0.05)--D(4096)E(0.05) & 5.46  & 4.29 & 45.6 & 4.88\\
\rownumber & C(128)L(0.05)--M(2)--D(4096)L(0.05)--D(4096)L(0.05)                                 & 5.99  & 8.62 & 105  & 5.18\\
\rownumber & C(128)L(0.05)--M(2)--D(4096)L(0.05)--D(4096)L(0.05)--D(4096)R                       & 5.34  & 5.18 & 61.1 & 5.86\\
\rownumber & C(128)L(0.05)--M(2)--D(4096)L(0.05)--D(4096)L(0.05)--D(4096)L(0.05)                 & 4.48  & 5.18 & 61.1 & 4.61\\
\rownumber & C(128)L(0.05)--M(2)--D(4096)L(0.05)--D(4096)L(0.05)--D(4096)L(0.05)--D(4096)L(0.05) & 4.23  & 4.28 & 44.9 & 4.86\\
\rownumber & C(256)E(0.05)--M(2)--D(4096)E(0.05)--D(4096)E(0.05)                                 & 6.46  & 8.18 & 93.0 & 5.05\\
\rownumber & C(256)E(0.05)--M(2)--D(4096)E(0.05)--D(4096)E(0.05)--D(4096)E(0.05)                 & 5.40  & 4.98 & 56.5 & 5.12\\
\rownumber & C(256)E(0.05)--M(2)--D(4096)E(0.05)--D(4096)E(0.05)--D(4096)E(0.05)--D(4096)E(0.05) & 5.98  & 4.08 & 41.4 & 5.28\\
\rownumber & C(256)L(0.05)--M(2)--D(4096)L(0.05)--D(4096)L(0.05)                                 & 6.14  & 7.97 & 93.6 & 5.29\\
\rownumber & C(256)L(0.05)--M(2)--D(4096)L(0.05)--D(4096)L(0.05)--D(4096)L(0.05)                 & 4.68  & 5.00 & 56.0 & 5.60\\
\rownumber & C(256)L(0.05)--M(2)--D(4096)L(0.05)--D(4096)L(0.05)--D(4096)L(0.05)--D(4096)L(0.05) & 4.32  & 4.10 & 41.7 & 4.94\\
\rownumber & C(256)S--M(2)--D(4096)S--D(4096)S--D(4096)S--D(4096)S                               & 11700 & 1400 & --- & --- \\
\rownumber & C(256)T--M(2)--D(4096)T--D(4096)T--D(4096)T--D(4096)T                               & 11800 & 49.5 & --- & --- \\
\hline
\multicolumn{6}{l}{\textbf{Notes.} Models trained for 20 epochs in batches of 64.}\\
\multicolumn{6}{l}{$^*$ Triangular2 learning rate policy ranging from $8 \times 10^{-6}$ -- $5 \times 10^{-3}$ with a complete cycle spanning 8 epochs.}\\
\multicolumn{6}{l}{$^{\dagger}$ Like LR1, but ranging from $10^{-5}$ -- $10^{-3}$.}\\
\multicolumn{6}{l}{$^{\ddagger}$ Constant learning rate of $10^{-5}$.}\\
\multicolumn{6}{l}{$^{\mathsection}$ Constant learning rate of $10^{-3}$.}\\
\multicolumn{2}{l}{$^a$ Dense layer with n nodes, D(n)}\\
\multicolumn{2}{l}{$^b$ ReLU activation}\\
\multicolumn{2}{l}{$^c$ Sigmoid activation}\\
\multicolumn{2}{l}{$^d$ tanh activation}\\
\multicolumn{2}{l}{$^e$ ELU activation E$(\alpha)$, with scaling parameter $\alpha$}\\
\multicolumn{2}{l}{$^f$ Leaky ReLU activation L$(m)$, with a slope of $m$ for $x < 0$}\\
\multicolumn{2}{l}{$^g$ Convolution1D layer with a kernel size of 3 and n nodes, C(n)}\\
\multicolumn{2}{l}{$^h$ MaxPooling1D layer M$(s)$, with a pooling size $s$}
\end{longtable}

\section{Data Set Size Considerations}
\label{app:datasetsize}

To briefly investigate the effect of data set size for our problem, we consider three models in addition to that presented in Section \ref{sec:methods} (``Main'').  
The additional models are trained on around 25\% of the total data set (614,208 training, 171,584 validation, 78,848 testing).  
Two models were trained according to the same learning rate policy as described in Section \ref{sec:methods}, for 187 and 500 epochs (``Sub1a'' and ``Sub1b'', respectively). 
The third model was trained according to the LR2 learning rate policy described in Appendix \ref{app:gridsearch} for 500 epochs (``Sub2'').
All other setup parameters (e.g., data normalization) match those described in Section \ref{sec:methods}.

Table \ref{tbl:size-comp} compares the normalized RMSE and denormalized $R^2$ test-set metrics over the high-resolution spectra, as well as the Bhattacharyya coefficients for the retrieved 1D marginalized posteriors.
Based on the differences between models Sub1a and Sub1b (which only differ in the number of epochs trained), it can be concluded that manually stopping training once the loss begins to negligibly change does not have a major effect on the model performance.  
Models Sub1b and Sub2 (which only differ in learning rate policies) illustrate the importance of selecting the learning rate policy.  
However, both of these effects are negligible compared to those of the data set size: the differences among models Sub1a, Sub1b, and Sub2 are smaller than the differences between model Main and Sub2 (the best-performing Sub model).  
While Main and Sub2 underwent similar numbers of total training steps, Main outperforms Sub2.  
These results motivate the generation of large, comprehensive data sets of spectra to train surrogate RT models, though further research into how data set size, number of inputs/outputs, and architecture complexity influence model performance is needed to inform optimal the data set sizes for future investigations.

\begin{table}[!hbt]
\caption{Model Comparison}
\label{tbl:size-comp}
\begin{center}
\begin{tabular*}{.56\textwidth}{l l l l l l}
 \toprule
Metric & Model & Min. & Median & Mean & Max. \\
\hline
Normalized RMSE & Main    & 0.00153 & 0.00224 & 0.00247 & 0.01040 \\
                & Sub1a   & 0.00271 & 0.00373 & 0.00407 & 0.01846 \\
                & Sub1b   & 0.00264 & 0.00365 & 0.00398 & 0.01679 \\
                & Sub2    & 0.00224 & 0.00331 & 0.00366 & 0.01721 \\
Denormalized $R^2$ & Main    & 0.99885 & 0.99993 & 0.99990 & 0.99997 \\
                   & Sub1a   & 0.99646 & 0.99980 & 0.99974 & 0.99991 \\
                   & Sub1b   & 0.99696 & 0.99981 & 0.99975 & 0.99992 \\
                   & Sub2    & 0.99740 & 0.99983 & 0.99977 & 0.99994 \\
Bhattacharyya coeff. & Main    & 0.9843 & 0.9948 & 0.9925 & 0.9972 \\
                     & Sub1a   & 0.9585 & 0.9858 & 0.9853 & 0.9991 \\
                     & Sub1b   & 0.8919 & 0.9933 & 0.9655 & 0.9976 \\
                     & Sub2    & 0.9783 & 0.9940 & 0.9918 & 0.9984 \\
\hline
\multicolumn{6}{p{0.95\linewidth}}{\textbf{Notes.}  See text for model descriptions.}
\end{tabular*}
\end{center}
\end{table}

\bibliography{MLRetrieval}{}
\bibliographystyle{aasjournal}



\end{document}